\documentclass{emulateapj}

\begin{document}


\title{Another Look at the EBS: A Stellar Debris Stream and a Possible
  Progenitor}

\author{Carl J. Grillmair}
\affil{Spitzer Science Center, 1200 E. California Blvd., Pasadena,  CA 91125}
\email{carl@ipac.caltech.edu}

\begin{abstract}

Using the Sloan Digital Sky Survey Data Release 7, we reexamine the
Eastern Banded Structure (EBS), a stellar debris stream first
discovered in Data Release 5 and more recently detected in velocity
space by Schlaufman et al. The visible portion of the stream is
$18\arcdeg$ long, lying roughly in the Galactic Anticenter direction
and extending from Hydra to Cancer. At an estimated distance of 9.7
kpc, the stream is $\approx 170$ pc across on the sky.  The curvature
of the stream implies a fairly eccentric box orbit that passes close
to both the Galactic center and to the sun, making it dynamically
distinct from the nearby Monoceros, Anticenter, and GD-1 streams.
Within the stream is a relatively strong, 2\arcdeg-wide concentration
of stars with a very similar color-magnitude distribution that we
designate Hydra I. Given its prominance within the stream and its
unusual morphology, we suggest that Hydra I is the last vestige of
the EBS's progenitor, possibly already unbound or in the final throes of
tidal dissolution. Though both Hydra I and the EBS have a relatively
high velocity dispersion, given the comparitively narrow width of the
stream and the high frequency of encounters with the bulge and massive
constituents of the disk that such an eccentric orbit would entail, we
suggest that the progenitor was likely a globular cluster, and that
both it and the stream have undergone significant heating over time.

\end{abstract}



\section{Introduction}

At least 14 stellar debris streams in the Galactic halo have now been
identified in photometric surveys (see \citet{grill2010} for a
review). A similar number of dynamically cold substructures have been
detected in velocity space \citep{helmi1999, smith2009,schlaufman2009,
  williams2011}.  Each of these streams is interesting as a partial
record of the accretion history of our Galaxy. However, and perhaps
more importantly, these streams can also serve as very sensitive
probes of the Galactic potential \citep{law2009,
  koposov2010}. Globular cluster streams are particularly important in
this respect as they are dynamically very cold \citep{combes1999,
  odenkirchen2009, willett2009}. A large sample of streams will
eventually enable us to constrain the distribution of dark matter in
the halo in a detailed and self-consistent manner. Enlarging the
sample of known streams will also increase the probability that we may
detect unmistakable signs of perturbations by dark matter sub-halos
\citep{mura99, carlberg2009, yoon2010}.

In this paper we reexamine the EBS first detected by
\citet{grill2006b} using the more complete coverage available in the
SDSS Data Release 7 (DR7, \citet{abazajian2009}). We briefly describe
our analysis in Section \ref{analysis}. We characterize the EBS and
and a possible progenitor in Section \ref{discussion} and we put
preliminary constraints on the orbit in Section \ref{orbit}. We make
concluding remarks Section \ref{conclusion}.

\section{Data Analysis} \label{analysis}

Data comprising $g,r,$ and $i$ photometry for $7 \times 10^7$ stars in
the region $108\arcdeg < \alpha < 270\arcdeg$ and $-4\arcdeg < \delta
< 65\arcdeg$ were extracted from the SDSS DR7 database using the SDSS
CasJobs query system. The data were analyzed using the matched filter
technique described by \citet{rock2002} and \citet{grill2009}.
Applied in the color-magnitude domain, the matched filter is a means
by which we can optimally differentiate between halo streams and the
foreground disk population.

In this paper we use filters based on the Padova database of
theoretical stellar isochrones \citep{marigo2008, girardi2010}. The
advantages of using theoretical isochrones include the ability to
explore a wider range of age and metallicity than is available among
the globular clusters within the SDSS footprint, as well as the
ability to extend the filters to very faint absolute magnitudes
(useful for examining very nearby structures).  These isochrones were
combined with the deep luminosity function of $\Omega$ Cen measured by
\citet{demarchi1999} to generate appropriate filters. We used all
stars with $15 < g < 22$, and we dereddened the SDSS photometry as a
function of position on the sky using the prescription of 
\citet{schlafly2011} applied to the $E(B - V)$ maps of
\citet{schleg98}. The field star distribution was sampled using
roughly half the Sloan survey area.  We applied the filters to the
entire survey area, and the resulting weighted star counts were summed
by location on the sky to produce two dimensional, filtered surface
density maps.

In Figure 1 we show the filtered star count distribution using a
filter based on an isochrone with Z = 0.0003 and an age of 13 Gyrs,
shifted in magnitude so as to provide optimal contrast for display
purposes for stellar populations at a distance of 9.7 kpc.  The surface
density map was generated by averaging the weights of each star based
on its distance from the $g - r$ and $g - i$ color-magnitude loci. The
surface densities have been binned to a pixel size of $0.1\arcdeg$ and
smoothed using a Gaussian kernel with $\sigma = 0.2\arcdeg$.

\begin{figure}
\epsscale{1.0}
\plotone{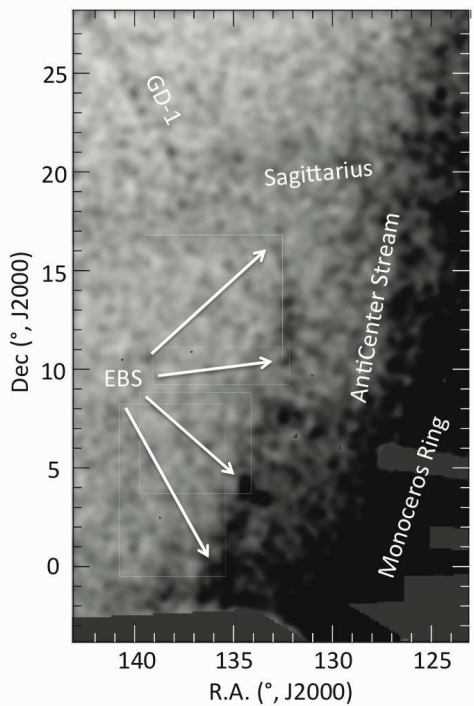}
\caption{Filtered surface density map of the southwest corner of the
  SDSS DR7 footprint. The stretch is logarithmic, and darker areas
  indicate higher surface densities. The map is the result of a filter
  based on a Padova isochrone with [Fe/H] = -1.8, an age of 13 Gyr,
  and shifted to a distance of 9.7 kpc. The results have been smoothed
  with a Gaussian kernel of width $0.2\arcdeg$, and no background
  subtraction has been applied.  Other known streams are indicated.}

\end{figure}

\section{Discussion} \label{discussion}

The region shown in Figure 1 is a complex of streams overlaid on a
rapidly rising population of foreground disk stars. Visible to varying
degrees (due to the (m-M)$_0$ = 14.83 magnitude shift of the filter)
are five well known features, namely the Sagittarius stream
\citep{belokurov2006a}, the Monoceros Ring \citep{newberg2002,
  yanny2003}, the Anticenter Stream (ACS, Grillmair 2006b), GD-1
\citep{grilld2006b}, and the ``Eastern Banded Structure'', or EBS
\citep{grill2006b}. While the EBS was only partly revealed in
\citet{grill2006b}'s DR5 analysis due to a large swath of missing
data, the additional coverage in DR7 allows us to trace the EBS for
some $18\arcdeg$ from the southern edge of the DR7 footprint in Hydra
to an indeterminate end in Cancer. The curvature of the stream takes
it to within $4\arcdeg$ of the similarly curved, southern end of the GD-1
stream, but there are clear discontinuities in position, distance, and
color-magnitude distribution that rule out any physical association
between them.

\citet{grill2009} used a significance test (the ``T-statistic'') that
measures the median contrast along its length between a putative
stream and the surrounding field.  The T-statistic for the EBS,
comparing with the field extending the length of the stream and
$15\arcdeg$ to the east, is shown in Figure 2.  The stream is clearly
not due to random fluctuations in the field; the filtered stream
signal is $\sim 28\times$ larger than the RMS measured using
identically sampled, neighboring field stars. A Gaussian that matches
the integrated, lateral profile of the stream has a
full-width-at-half-maximum of 1.0\arcdeg. At a distance of 9.7 kpc
(see below) this corresponds to a spatial extent perpendicular to our
line of sight of 170 pc. This is roughly twice as broad as typical,
presumed globular cluster streams \citep{odenkirchen2003,
  belokurov2006a,grillj2006, grilld2006a, grilld2006b, grill2009}, but
considerably narrower than the $>1$ kpc widths associated with
presumed dwarf galaxy streams \citep{maje2003,mart2004,grill2006a,
  grill2006b, belokurov2006b, belokurov2007, grill2009}. In the
absence of heating effects due to a rather eccentric orbit (see
below), this would suggest a progenitor that was significantly more
massive than the globular clusters believed to be responsible for the
currently known cold streams.

\begin{figure}
\epsscale{1.2}
\plotone{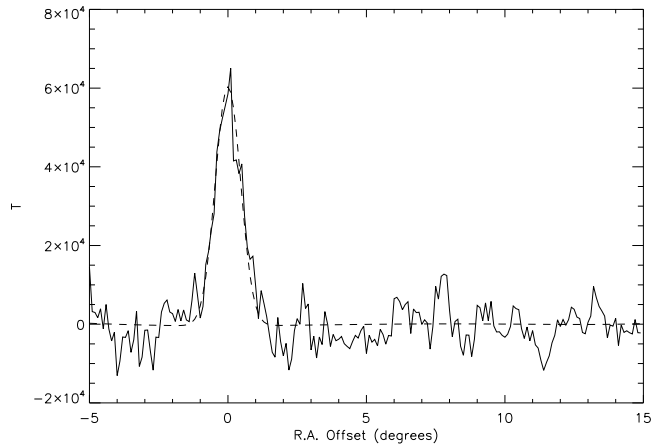}
\caption{The ``T'' statistic \citep{grill2009}, showing the
  background-subtracted, median filtered signal over five,
  3\arcdeg-long segments, integrated over a width of one degree, as a
  function of lateral offset from the stream.  The peak value is 28 times
  larger than the RMS measured for the identically sampled region
  between 2 and $15\arcdeg$ east of the stream, indicating a very low
  probability that the stream could be due to random fluctuations in
  the field.  The dashed line shows a Gaussian with a FWHM of
  1.0\arcdeg, which we take as a measure of the average breadth of the
  stream.}

\end{figure}

Figure 3 shows color-magnitude distributions (CMDs), dereddened as
prescribed by \citet{schlafly2011} using the $E(B - V)$ maps of
\citet{schleg98}, of stars lying within $1\arcdeg$ of the centerline
of the EBS, after subtraction of the CMD of stars lying between
$2.4\arcdeg$ and $4.0\arcdeg$ both east and west of the
stream. Isochrones with Z = 0.0003 ([Fe/H] $\approx -1.8$) and an age
of 13 Gyrs evidently match the turn-off and main sequence colors
reasonably well. We infer that the progenitor of the stream was old
and metal poor.

\begin{figure}
\epsscale{1.2}
\plotone{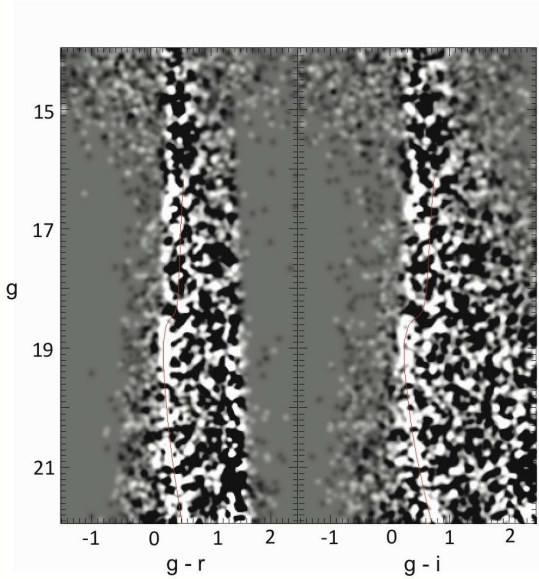}
\caption{Hess diagrams of the stars lying within $1\arcdeg$ of the
  centerline of the EBS. Padova isochrones with [Fe/H] =-1.8, age 13
  Gyrs, and shifted to a distance of 9.7 kpc are over-plotted. Lighter
  areas indicate higher surface densities.}

\end{figure}

Following \citet{grilld2006b} we shift the main sequence used to
construct our filter both brightward and faintward to estimate the stream's
distance. To avoid issues related to a possible difference in age
between our adopted isochrone and the stream stars, we use only the
portion of the filter with $19.5 < g < 22.0$, where the bright cutoff
is 0.3 mag below the main sequence turn off. We find that the strength
of the southern half of the stream peaks at a distance modulus of
$14.9 \pm 0.2$ mag, while the northern half of the stream peaks at
$14.8 \pm 0.3$ mag.  This puts the southern end of the stream at a
sun-centric distance of $9.7 \pm 0.9$ kpc, while the northern end is
at $9.4 \pm 1.4$ kpc. The portion of the stream visible in Figure 1 is
evidently almost perpendicular to our line of sight.

Integrating the background-subtracted, unfiltered counts of stars
within $3\sigma$ of the Z = 0.0003 isochrone along the length of the
stream and over a width of $1.5\arcdeg$ we find the total number of
stars in the discernible stream to be $530 \pm 230$. The large
uncertainty simply reflects the Poisson statistics of the very high
background, ($\approx 33,000$ field stars in the same region of color
and configuration space). Figure 4 shows a background-subtracted,
longitudonal profile of the filtered star counts, normalized to yield
an integrated total of 530 stars.  For stars with $g < 22$ and a
stream width of 1.0\arcdeg, the average surface density is $30 \pm 13$
stars deg$^{-2}$, with a peak of over 100 stars deg$^{-2}$. Like the
Pal 5 and GD-1 streams, the EBS profile shows interesting peaks and
troughs, fairly regularly spaced with a separation of $4.0\arcdeg \pm
0.2\arcdeg$.  While the uncertainties are large, the clumps and gaps
that give rise to these features appear quite obvious in Figure 1. The
regular spacing may suggest an origin tied to the orbit of the stream,
perhaps a result of episodic stripping (e.g. major stripping pulses at
the perigalacticon of a highly eccentric orbit \citep{grillmair1992,
  johnston1995}). Alternatively, the undulations may be due to
scattering by encounters with massive objects in the disk or halo
\citep{mura99, yoon2010}.

\begin{figure}
\epsscale{1.2}
\plotone{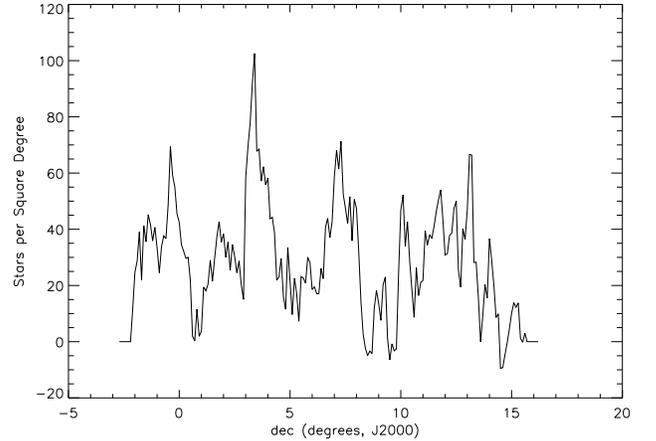}
\caption{The longitudonal profile of the filtered star counts,
  boxcar smoothed with a width of 0.5\arcdeg. The
  profile is measured over a stream width of $1.5\arcdeg$ and is
  background subtracted using the distribution of filtered star counts
  between 2 and $4\arcdeg$ east and west of the stream. The profile has
  been normalized to yield an integrated total of 530 stars to $g =
  22$. }

\end{figure}

\citet{schlaufman2009} recently detected a number of cold halo
substructures in velocity space (``ECHOS'') using SEGUE data. With a
plate center at (R.A., dec) = ($132.6\arcdeg$, $6.1\arcdeg$), their
B-7/PCI-8/PCII-20 detection overlaps the EBS (Figure 5) and the
estimated distance of 10 kpc is almost identical to what we find for
the stream. We have examined this region using isochrone filters with
[Fe/H] ranging from -2.2 to 0.0 and find no evidence for other
cold substructures at this distance. We conclude that
B-7/PCI-8/PCII-20 is most likely sampling stars in the EBS stream.

\citet{schlaufman2009} find a mean radial velocity for
B-7/PCI-8/PCII-20 of 71 km s$^{-1}$, and a dispersion of 13 km
s$^{-1}$.  This dispersion is significantly larger than that measured
for known and presumed globular cluster streams
\citep{odenkirchen2009, willett2009, koposov2010}. On the other hand,
it is quite similar to measurements of presumed dwarf galaxy streams
\citep{grill2008, carlin2010, newberg2010}.  Combining the width of
the stream with its large apparent velocity dispersion, we might infer
that the stream's progenitor was substantially more massive than Pal 5
or the globular clusters that produced GD-1, Acheron, Cocytos, or
Lethe. On the other hand, the stream is neither as broad nor as
populous as streams associated with classical dwarf galaxies like
Sagittarius or the progenitor of the Orphan Stream. The high velocity
dispersion may be partly due to non-EBS stars in the sample, or it may
be due to heating of the progenitor by disk or perigalactic shocking
prior to the stripping of these stars. It may also be due to
significant heating by encounters with either dark matter subhalos
\citep{carlberg2009} or massive structures (e.g. giant molecular
clouds) in the disk. Another possibility is that the EBS may be the
remnant of an ultrafaint dwarf galaxy \citep{willman2005, grill2006a,
  zucker2006a, zucker2006b}, though as we discuss below, it is
difficult to imagine how such an object could have retained a dark
matter envelope for any length of time.

\subsection{Hydra I: A Disrupting Progenitor?} \label{hydra1}

Figure 5 shows an expanded, lower-contrast view of the southern
portion of the EBS. With a contrast maximum at the same distance
($\approx 9.7$ kpc) as the EBS is an interesting and relatively
compact feature at [R.A., dec] $\approx [133.9\arcdeg, 3.6\arcdeg$].
The object contains $\approx 300$ stars to $g = 22$, has a filtered
star count density higher than any visible portion of the EBS, and appears
quite distinct within the stream.  There are several background galaxy
clusters identified within $1\arcdeg$ of this position, but
examination of the identically filtered SDSS DR7 galaxy catalog shows
no significant galaxy concentration of similar size or shape. Two
lesser peaks are apparent some $4\arcdeg$ north and south of this
object (see also Figure 4), but we focus on this object because it is
the most prominant and populous concentration, both to the eye and in
the longitudonal profile.

\begin{figure}
\epsscale{1.0}
\plotone{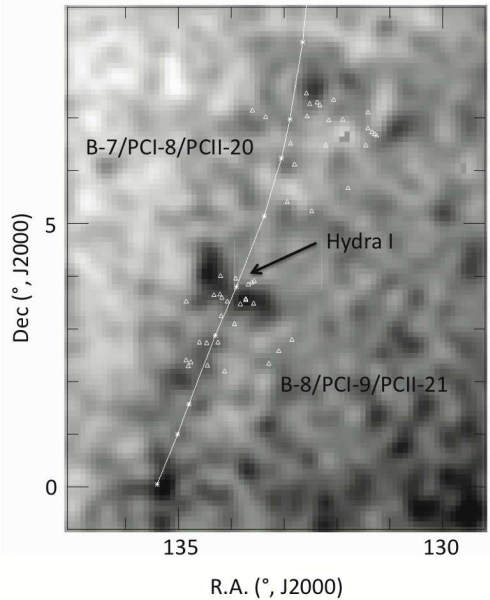}
\caption{An expanded, lower-contrast view of the southern portion of
  the EBS, showing the positions of the stars making up
  \citet{schlaufman2009}'s B-7/PCI-8/PCII-20 and B-8/PCI-9/PCII-21
  ECHOS detections (Schlaufman, private communication). The asterisks
  connected by lines show the normal points used to define the path of
  the stream for fitting purposes.}

\end{figure}

Given its apparent location at the same distance as the stream and
approximately centered within it, we infer that the feature is
physically associated with the EBS and we designate it Hydra I. The
feature appears somewhat amorphous, with two primary concentrations
extending to the north and west, respectively, for a total extent of
$\approx 2\arcdeg$.  At 9.7 kpc this corresponds to a spatial extent
of about $\sim 350$ pc, which is far larger than any known globular
cluster. Using filter shifts to estimate relative distances, we find
that the maximum filtered surface densities in the two lobes of Hydra
I occur within 0.1 magnitudes of one another, indicating that Hydra I
is not significantly extended along our line of sight. At 9.7 kpc, 0.1
magnitudes corresponds to a difference in distance of 500 pc. To
within the uncertainties, this is identical to the lateral extent of the
object. 

How real is the apparent, double-lobed morphology of Hydra I? Could
the northeastern lobe be simply a chance consequence of Poisson
statistics? The surface density profile of stars with $19 < g < 23$ in
the stronger, western lobe and lying within 0.2 magnitudes of the
Z=0.0003 ischrone in $g -i$ is shown in Figure 6. Fitting an
elliptical $1/r$ model to the star counts, we find ellipticity
$\epsilon = 0.45 \pm 0.05$, with $\theta = 90 \pm 10\arcdeg$ (measured
north through east), a total population out to $1\arcdeg$ of $300 \pm
10$ stars, and an overall $\chi^2$ of 1.2. Following
\citet{martin2008}, we then determine the fractional R.M.S. deviation
$\sigma_{sc}/$total of the data compared to the model. We generate
1000 Poisson realizations of the field out to $1\arcdeg$ and examine
the distribution of ($\sigma_{sc}$/total)$^2$. The peak of the
distribution differs from 0 at the $9\sigma$ level, indicating that
the northeastern lobe is unlikely to be a statistical departure from
the model. We can only speculate as to the relationship between the
two lobes at this point.  However, being part of a tidal stream, it
seems reasonable to suppose that the two lobes are unbound, comoving
tidal remnants. Deeper imaging of Hydra I is currently being acquired
and will be the subject of a future contribution.

\begin{figure}
\epsscale{1.2}
\plotone{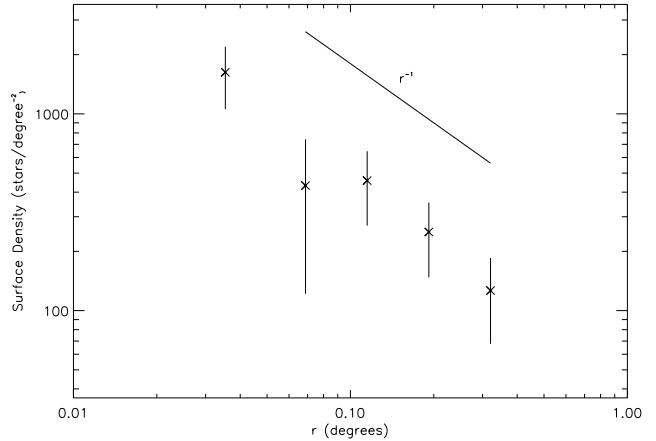}
\caption{The surface density profile of stars in Hydra I, measured
  with respect to the center of the western lobe at (R.A., dec) =
  ($133.450\arcdeg, 3.465\arcdeg$). Only stars with $19.0 < g <
  23.0$ and lying within 0.2 magnitudes of the Z=0.0003 $g - i$
  isochrone are counted. The background level was measured using an
  annulus with $2.0\arcdeg < r < 3.0\arcdeg$. 
}

\end{figure}

\citet{schlaufman2009} detected a velocity overdensity among metal
poor stars at [R.A., dec] = [$134\arcdeg, 3.2\arcdeg$] that they
attributed to the Monoceros ring. Figure 5 shows that the stars making
up their B-8/PCI-9/PCII-21 detection clearly sample the position of
Hydra I, as well as a portion of the EBS extending to the south (which
would be expected to have a nearly identical velocity). Their
estimated distance of 9.7 kpc is identical with the values we find
above. Filtering this region of DR7 using isochrones with
metallicities spanning the range $-2.2 < $ [Fe/H]$ < 0.0$ reveals no
other significant structures at this position. We conclude that
\citet{schlaufman2009} actually sampled the stars in Hydra I and the
EBS. They find a mean velocity for B-8/PCI-9/PCII-21 of 85 km s$^{-1}$
and a dispersion of 14.9 km s$^{-1}$.  The velocity dispersion is
again quite high compared with globular clusters or ultrafaint
galaxies, suggesting that Hydra I was either quite massive or has
been significantly heated over time. Based on velocity dispersion,
\citet{schlaufman2011} have suggested that the progenitors of their
ECHOS were dwarf spheroidal galaxies. However, based on the the orbit
constraints below, the high velocity dispersions of the EBS and Hydra I
may be due to significant heating by encounters with the dark matter
subhalos \citep{carlberg2009} or massive star clusters or molecular
clouds in the disk.

\subsection{Constraints on the Orbit} \label{orbit}

Mindful of the fact that tidal streams do not precisely trace the
orbits of their progenitors \citep{odenkirchen2009, eyre2009, eyre2011}, we
nevertheless estimate the orbit of the {\it stream} stars to determine
whether the EBS might be related to any of the other streams in Figure
1. We do this using the Galactic model of \citet{allen91} to compute
trial orbit integrations, and matching these orbit integrations with
the measured positions, distances, and velocities of the stream in a
least-squares sense. We integrate orbits over a grid of possible
radial velocities and proper motions, using the IDL AMOEBA downhill
simplex procedure to find the minimum $\chi^2$ at each grid point.
The grid points are separated by 1 km s$^{-1}$ in radial velocity, and
0.02 mas yr$^{-1}$ in each component of proper motion. 

We fit to 17 normal points chosen to lie along the estimated
centerline of the stream.  We use a solar Galactocentric distance of
8.5 kpc, and stream distances and velocities as given above.  We adopt
velocity uncertainties for the measurements at B-7/PCI-8/PCII-20 and
B-8/PCI-9/PCII-21 of 3 km s$^{-1}$ (Schlaufman, private
communication), positional uncertainties of $0.2\arcdeg$, and distance
uncertainties as given above. We have attempted to measure the proper
motions along the stream using the proper motions provided in DR7
\citep{munn2004, munn2008}.  Unfortunately, the relatively large
uncertainties and severe contamination by field stars conspire to wash
out any obvious signal due to the stream. We consequently leave the
proper motions as free parameters in the fit. Tighter constrains on
the orbit will have to await the acqusition of more accurate proper
motions with Gaia or the Large Synoptic Survey Telescope.

Figure 7 shows the normal points used to fit the postions along the
EBS stream. Also shown are projections of the best-fit orbits on the
SDSS DR7 footprint. Both the prograde and retrograde orbit models
predict that the stream passes between 1 and 2 kpc of the current
position of the sun, near the north Galactic pole in projection.  At
this distance, the stream would be more than $5\arcdeg$ across, and
moving between three and four times faster than at apogalacticon. If
the stellar stream extends along these portions of the orbits, and if
the number of stars stripped from the progenitor per unit time was
roughly constant over the lifetime of the stream, then we would expect
the surface density near the north Galactic pole to be $\sim 15 - 20$
times less than it is in Figure 1. Since we expect the rate of tidal
stripping to accelerate as the mass of the progenitor is diminished
over time, the relative number of stream stars we might expect to see
far from the progenitor would be reduced further still. If the orbit
is retrograde, then the star count signal of the EBS would likely be
buried within the much more populous Sagittarius stream. If there are
EBS stars passing near the sun, they will more easily be found in
velocity and proper motion surveys (e.g. RAVE, Gaia, LSST).

\begin{figure}
\epsscale{1.3}
\plotone{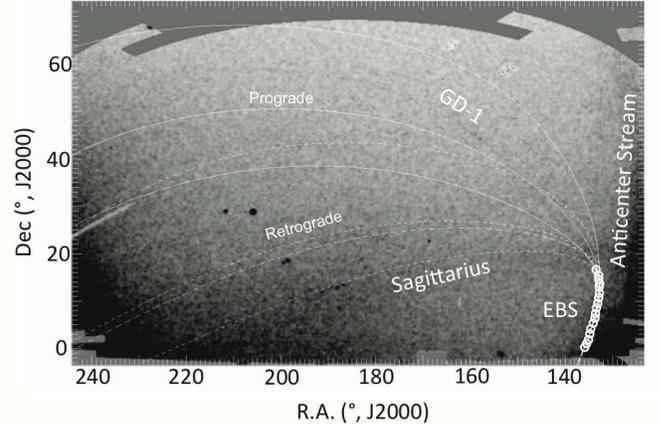}
\caption{Best-fit orbits for the EBS projected onto the SDSS DR7
  footprint. Open circles indicate the normal points used to trace the
  stream and constrain the fit. The solid lines show the best-fitting
  prograde orbit, along with the 90\% confidence limits. The
  short-dashed lines similarly show the best-fit and 90\% limits on the
  retrograde orbit. The long-dashed line shows the best fit to the data
  if no velocity constraints are imposed.}

\end{figure}

Figure 8 shows orbit integrations that correspond to the best-fit
parameters for both prograde and retrograde models. While the formal
$\chi^2$ for the retrograde model is 15\% less than that for the
prograde model, the disagreement between the predicted velocity
gradient along the stream (3.1 km s$^{-1}$ deg$^{-1}$) and the
measurements at B-7/PCI-8/PCII-20 and B-8/PCI-9/PCII-21 (-4.8 km
s$^{-1}$ deg$^{-1}$) is somewhat larger (and of the opposite sign)
than for the prograde model (-10.7 km s$^{-1}$ deg$^{-1}$).
Additional velocity measurements at different positions along the
stream and/or proper motion measurements will be required to resolve the
ambiguity. The proper motions predicted at the position of Hydra I are
$\mu_\alpha \cos{\delta} = -0.15$ mas yr$^{-1}$, $\mu_\delta = -2.67$ mas
  yr$^{-1}$ for the prograde orbit, and $\mu_\alpha \cos{\delta} =
  +0.65$ mas yr$^{-1}$, $\mu_\delta = -5.08$ mas yr$^{-1}$ for the
retrograde orbit. These values are of the same magnitude or less
than the typical DR7 uncertainties ($\approx 4$ mas yr${-1}$), so it is
perhaps not surprizing that we have been unable to identify a clear
stream signature in the proper motion data.

\begin{figure}
\epsscale{1.8}
\plotone{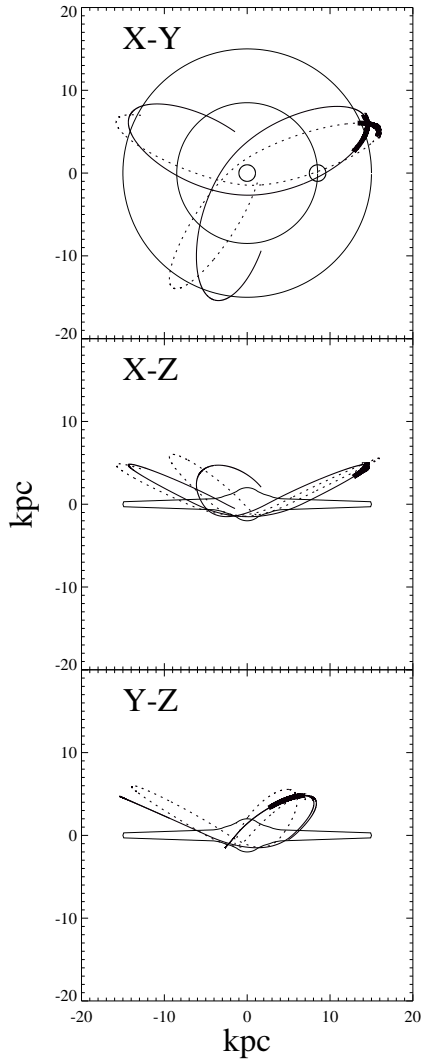}
\caption{Best-fit orbit projections for the EBS in X, Y, and Z Galactic
  coordinates. The heavy lines show the portions of the orbit with $
  -2\arcdeg < \delta < 16\arcdeg$. The thin solid curve shows the
  best-fit prograde orbit, while the dotted curve shows the retrograde
  orbit that best fits all the data. The sun's location at (X,Y,Z) =
  (8.5,0,0) kpc is indicated.}

\end{figure}

The prograde orbit model predicts apogalacticon R$_A = 16.5 \pm 0.1$
kpc, perigalacticon R$_P =3.0^{+0.7}_{-0.3}$ kpc, eccentricity
$\epsilon = 0.69^{+0.02}_{-0.05}$, and inclination $i = 17\arcdeg \pm
0.4\arcdeg$, where the uncertainties correspond to the 90\% joint
confidence interval. For the retrograde orbit, the parameters are R$_A
= 17.7 \pm 0.4$ kpc, R$_P = 1.8^{+0.4}_{-0.2}$ kpc, $\epsilon =
0.8^{+0.02}_{-0.03}$, and $i = 19.9\arcdeg \pm 0.7\arcdeg$. In either
case, the EBS appears to be on a fairly eccentric orbit. This
eccentricity is forced entirely by the curvature of the stream at its
northern end. If we give zero weight to the B-7/PCI-8/PCII-20 and
B-8/PCI-9/PCII-21 velocities, the best-fitting orbit still predicts an
eccentricity of $\epsilon \approx 0.8$. The stream is evidently not
associated with either the ACS or the Monoceros Ring, both of which
have been determined to be on very nearly circular orbits
\citep{penarrubia2005, grill2008}.  The combination of inclination and
eccentricity takes the stream into the inner, non-spherical part of
the Galactic potential, where no component of angular momentum is
conserved and we see an interesting box orbit as a
consequence. Frequent, oblique passages through the disk would
presumably increase the potential for encounters with massive
structures such as stellar clusters or giant molecular clouds and may be
partly or wholly responsible for the high velocity dispersion observed
in the Hydra I and the EBS stream.

\section{Conclusion} \label{conclusion}

The EBS stream adds to the growing list of halo streams that can be
mapped over a sufficient extent that, with suitable follow-up
observations, they could be used as probes of the Galactic
potential. A preliminary orbit estimate shows that the EBS is
unrelated to either the Anticenter or Monoceros streams.  The somewhat
intermediate breadth of the stream together with its relatively high
velocity dispersion suggests the possibility that the progenitor could
have been more massive than the globular clusters thought to be
responsible for the half dozen very cold streams discovered in the
SDSS footprint to date. However, if the progenitor had been a dark
matter dominated dwarf galaxy, it would be difficult to understand how
it could have held onto it's dark matter envelope for any length of
time in such a confined and eccentric orbit.  On the other hand, this
very orbit may have subjected both the progenitor and the stream to
significant heating through encounters with massive structures in the
disk.

If Hydra I is indeed the progenitor of the EBS, then it is only the
second probably unbound progenitor to be associated with a tidal
stream. A more detailed examination of the structure and stellar
kinematics in this remnant may shed new light on the end stage of
tidal disruption.  Though contamination by field stars is high, Hydra
I may be particularly attractive in this respect as it is four times
closer to us than Bootes III \citep{grill2009}.

Refinement of the orbit will require radial velocity and proper motion
measurements of carefully selected stars along the length of the
stream. Given the very low surface density of stream stars and very
high field star contamination, this will necessarily be an ongoing
task. In this respect, the EBS may be particularly well situated for
follow-up by the upcoming spectroscopic LAMOST survey. Gaia and LSST proper
motion measurements may also help us to refine the orbit and perhaps
trace the stream over a much longer arc.

\acknowledgments

The author is grateful to an anonymous referee for 
constructive and insightful comments. Thanks also go to Kevin
Schlaufman for providing the positions of ECHOS member stars. 
Funding for the creation and distribution of the SDSS Archive has been
provided by the Alfred P. Sloan Foundation, the Participating
Institutions, the National Aeronautics and Space Administration, the
National Science Foundation, the U.S. Department of Energy, the
Japanese Monbukagakusho, and the Max Planck Society.

{\it Facilities:} \facility{Sloan}.

\end{document}